# ON THE HADRONIC PRODUCTIONS OF $B_c$ AND $B_c^*$ MESONS

Chao-Hsi CHANG

CCAST (World Laboratory), P.O.Box 8730, Beijing 100080, China
Institute of Theoretical Physics, Academia Sinica, P.O. Box 2735, Beijing 100080, China
International Center for Theoretical Physics, P.O. Box 586, I-34100 Trieste, Italy
E-mail: ZHANGZX@itp.ac.cn

ABSTRACT

With respect to the two theoretical approaches for estimating the hadronic productions of the mesons $B_c$ and $B_c^*$: the full $\alpha_s^4$ calculation and the fragmentation approximation, with a thorough comparative study a quite remarkable conclusion may be drawn. In hadronic collisions the condition for the applicability of the fragmentation approximation is the transverse momentum $P_T$ of the produced $B_c$ and $B_c^*$ being much greater than the meson masses, and the higher the energy of the relevant subprocess the higher the $P_T$ is requested. The mechanisms for the productions are also clarified. In order to show the resultant differences of the two approaches, newly introduced distribution of the fraction of the $B_c$ (or $B_c^*$) meson energy (a measurable observable) in the center of mass of the subprocess is emphasized.

The reason why we would like to talk about the subject here is three-fold: first of all, the problem is interesting itself; secondly, there were some obscurities due to a few incorrect numerical results and misleadings, and now the situation has been clarified; the third, some fresh results have been achieved.

The heavy flavored $\bar{b}c$ meson states have attracted considerable interest due to their interesting properties ( the spectroscopy of $\bar{b}c$ meson states[1]; the weak decays[2] and the productions[3-18] etc ). The experimental search for these mesons is now under way at high energy colliders such as the LEP $e^+e^-$ collider (at $Z^0$ resonance) and the Tevatron $\bar{p}p$ collider (at the full energy 1.8 TeV). Various experimental results are expected to come out soon thus to estimate the production cross sections as precisely as possible has become a desirous theoretical task. Fortunately the mesons are weakly bound states and relatively simple, their hadronization is calculable in the framework of perturbative QCD (PQCD) in terms of the wave function obtained by potential model. At LEP being relatively simple, the productions of the pseudoscalar ground state $B_c$ and the vector meson state $B_c^*$ are dominated by $Z^0$ decay into a $b\bar{b}$ pair, followed by the 'fragmentation' of a $\bar{b}$ quark into the $B_c$ or $B_c^*$ meson[3,4,5], whereas as first pointed out in Ref.[6], the energetic hadronic productions are much more complicated even only to the lowerst order of PQCD in $\alpha_s^4$ and dominated by the subprocess of 'gluon-gluon fusion' $(gg \to B_c(B_c^*)b\bar{c})$. In practice, a possible and alternative way to calculate the hadronic productions is to apply the fragmentation approximation. With the approximation the calculation can then be considerably simplified, as indicated in Ref.[7]. Since then one interesting question is addressed, how well and/or in which kinematic region for the hadronic productions of the $B_c$ and $B_c^*$ mesons the fragmentation approximation works. Therefore several groups, based on the same consideration, have recalculated the productions. Namely since Ref.[6] presented the numerical results for the hadronic productions first, recalculations have been completed and distributed[8-15]. The calculations for the hadronic productions of the mesons $B_c$ and $B_c^*$ to order $\alpha_s^4$ in PQCD involve very complicated numerical calculations. At the first stage, not all the calculations were in agreement: in Ref.[8] an order of magnitude larger result than Ref.[6] is claimed; in Ref.[9] again a result larger than Ref.[6] but smaller than Ref.[8] is obtained; a result similar to Ref.[9] is found by Masetti and Sartogo[10]. But more recently, it is found by the same authors of Ref.[9] that a color factor $\frac{1}{3}$ had been overlooked in their previous work and after correcting this factor, a result in agreement with Ref.[6] has been achieved[11]. In Ref.[12] similar numerical results to Ref.[6] have also independently been obtained. In Ref.[13] different numerical results have been presented, whereas it is difficult to compare directly with others due to different parton distribution functions and different energy scale being adopted in the calculation. Thus in Refs.[14,15], the authors of Ref.[6] as well as their collaborators have re-studied the productions extensively with the updated parton distribution functions[19] and made thorough comparisons with others' [11-13]. Now at least of the four groups[6,14,15,11,12,13], the numerical results for the cross sections of the subprocess are consistent each others within the uncertainties of the numerical calculations. One may now be confident that the numerical results of the full order $\alpha_s^4$ PQCD calculations are in agreement. Thus I will discuss the problem based on the confident results.

From experience, it is naively believed that, when the transverse momentum $P_T$ of the meson $B_c$ or $B_c^*$ is large, the hadronic productions of the mesons are dominated by jet fragmentation, so the fragmentation approximation may become very valid. However one should carefully examine the approximation not only because it will potentially be used as a further test of the factorization theorem of QCD but also because of special interest in estimating the productions as reliable as possible i.e. as mentioned above to predict the productions precisely so as to guide the discovery of the meson $B_c$ in experiments. To pursue the goal, subsequent comparisons between the PQCD full $\alpha_s^4$ calculation and the fragmentation approximation were made by several groups[6,14,15,11,13]. Though all the results were comparable, different conclusions still were drawn due to different observables in different kenetics regions being emphasized. I will talk about the aspects in detail below.

Based on PQCD to calculate the hadronic productions, the full $\alpha_s^4$ order calculation is illustrated in Fig.1a with Fig.1b and the fragmentation approach in Fig.1a and Fig.1c.

As pointed out above, it is very interesting to investigate the appoaches. Since the $P_T$ distribution decreases very rapidly as $P_T$ increases, one would lose a lot of statistics if one had only considered large $P_T$ events. Thus the comparitively low $P_T$ events should also be considered carefully, as long as they may escape from the experimentally proper cuts. In Ref.[11] by examining the production ratio for $B_c^*$ to $B_c$, it is claimed that the fragmentation approximation breaks down even for very large $P_T$, whereas in Ref.[13] by investigating the $P_T$ distribution of the $B_c$ meson, it is claimed that the fragmentation approximation works well if $P_T$ exceeds about $5-10$ GeV. Whereas we investigated this problem quite early[14] and re-examined it more carefully thus a more completed feature has been achieved[15].

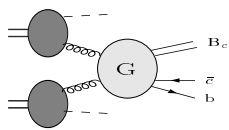

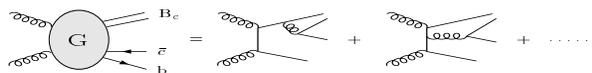

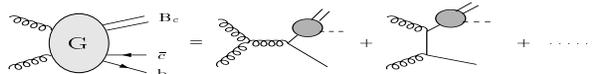

Figure 1: **The factorizations for the productions**

To clarify the problem, let us focus the discussions on the subprocess. To the lowest order $(\alpha_s^4)$, there are 36 Feynman diagrams responsible for the dominant gluon fusion subprocess $g(k_1) + g(k_2) \to B_c(p) + b(q_2) + \bar{c}(q_1)$, where $k_1$, $k_2$, $p$, $q_1$, and $q_2$ are the respective momenta. As pointed out in Refs.[6,14], of the 36 Feynman diagrams one may split them into five independent "groups" according to the color structure and each of them is gauge invariance for QCD. Furthermore, the contributions from the kinematic region where certain factors of the amplitude for the process are nearly singular; i.e., some of the internal quark lines and gluon lines in the Feynman diagrams are close to mass-shell are substantial even dominant, especially when the c.m.s. energy of the subprocess, $\sqrt{\hat{s}}$, is much larger than the heavy quark mass. Specifically, here for the concerned subprocess and in a special chosen gauge where large number cancellation due to gauge invariance does not happen at all, the possible singularities may arise from the inverse power(s) of the following factors, or their products:

$$q_i \cdot k_j, \qquad p \cdot k_j, \qquad (\alpha_i p + q_i)^2, \qquad \text{and} \qquad (k_j - \alpha_1 p - q_1)^2, \tag{1}$$

which corresponds to the denominators of the quark and gluon propagators appearing in the amplitude and here $i, j = 1, 2$ and $\alpha_{1,2} = \dfrac{m_{c,b}}{(m_c + m_b)}$ is the ratio of quark masses. The singularities result in the cross section, for the subprocess and upto the lowest twist contributions, proportional to $\dfrac{1}{\hat{s}} \dfrac{f_{B_c}^2}{M_{B_c}^2}$ (where $f_{B_c}$ is the $B_c$ decay constant) and with some logarithmic correction terms such as $\ln(\hat{s}/M_{B_c}^2)$ being involved. When the $P_T$ of the $B_c$ meson is large, only $(\alpha_i p + q_i)^2$ can still be small ($\sim m_i^2$), therefore the fragmentation functions can then be extracted from the most singular part containing the inverse powers of this factor in the square of the amplitude. It then follows that in the large $P_T$ region the subprocess is dominated by the fragmentation. *However, when the $P_T$ of the $B_c$ meson is small, the produced $B_c$, as well as the b and the $\bar{c}$ quarks, can be soft or collinear with the beam.* In this region the amplitude is highly singular because *two or more* of the internal quarks or gluons in certain Feynman diagrams can *simultaneously* be nearly on-mass-shell. Although this region is a smaller part of the phase space, these nearly singular Feynman diagrams make a substantial contribution and dominate to the cross section in the region. In fact, in the square of the amplitude we can isolate all the terms which contribute to the lowest twist cross sections using singularity power counting rules[20]. When $\sqrt{\hat{s}} \gg M_{B_c}$ the lowest twist contributions dominate, while the higher twist contributions are suppressed by a factor $m^2/\hat{s}$ at least and small. We can decompose the terms which contribute to the lowest twist cross sections into three components: One is due to the fragmentation contribution, which dominates the large $P_T$ region, and the second is due to the non-fragmentation, which comes from the other singular parts as discussed above (those in which *two or more* quarks or gluons are nearly on-mass-shell). The non-fragmentation dominates in the smaller $P_T$ region. As for the third, the regular parts, they contribute the smooth "background" only and small, thus they are not so interesting. The contributions from the first two components are quite clearly distinguishable in the $P_T$ distribution of the subprocess, particularly at large $\sqrt{\hat{s}}$.

In order to show the factor quantatively, the cross sections for the dominant subprocess $gg \to B_c(B_c^*) + \cdots$ versus $P_T$ (without any $P_T$ cut) are calculated by the two approaches and at four energies for the colliding gluons: 30GeV, 60GeV, 100GeV and 200GeV. The results are plotted in Fig. 2 (Fig. 2a-2d). Here the solid lines indicate those for the full QCD $\alpha_s^4$ calculations; the dotted lines indicate those for the fragmentation approximation. Note that in the calculations here the values of the parameters $\alpha_s = 0.2$, $m_b = 4.9$ GeV, $m_c = 1.5$ GeV, $M_{B_c} = 6.4$ GeV and $f_{B_c} = 480$ MeV are taken, moreover in the fragmentation calculations, in order to reduce the error caused by the phase space integrations, we directly used the squared matrix elements from which the fragmentation functions are derived, rather than the fragmentation functions themselves.

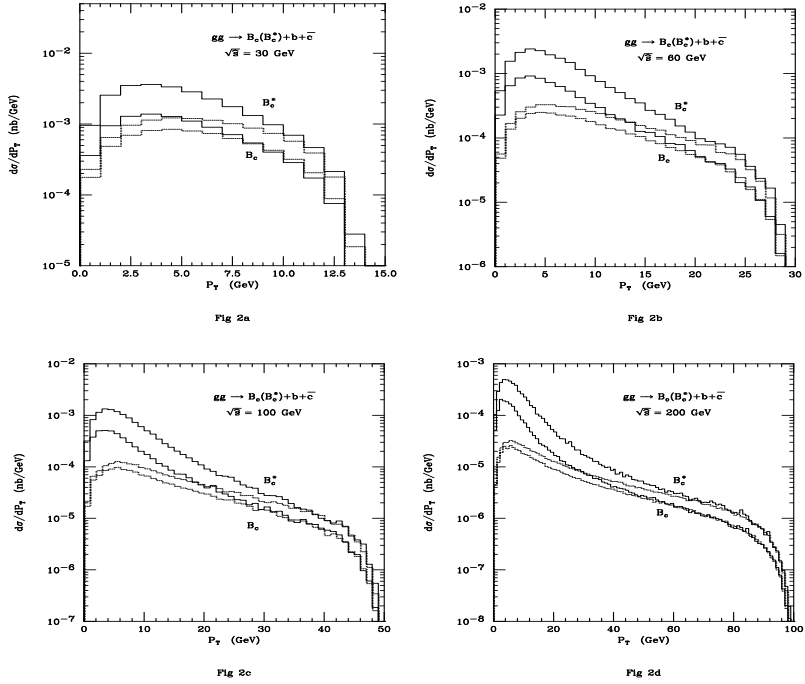

Figure 2: **The cross sections for various colliding energies**

Let us take the result at a very high colliding enery $\sqrt{\hat{s}} = 200$ GeV as an example to compare the full $\alpha_s^4$ calculation and the fragmentation approximation. It is easy to see from Fig. 2d that when $P_T$ is larger than about 30 GeV for the $B_c$ and about 40 GeV for the $B_c^*$, the fragmentation approximation is close to the full $\alpha_s^4$ calculation. However, when the value of the $P_T$ becomes smaller and smaller than about 30 GeV for the $B_c$ and about 40 GeV for the $B_c^*$, the deviation between the fragmentation approximation and the full calculation becomes larger and larger. The non-fragmentation component clearly starts to dominate the productions at certain critical value of $P_T$, which certainly is much larger than the $B_c$ meson mass. If comparing all the four figures Fig. 2a-2d it is easy to see that the critical value of $P_T$ increases slowly with $\hat{s}$ is increasing, that indicates there is an additional enhancement maybe due to logarithmic terms such as $\ln \hat{s}/m^2$ in the non-fragmentation component compared to the fragmentation component. When $\sqrt{\hat{s}}$ is not very large this two component decomposition is less distinct, since the higher twist terms cannot be ignored. In summary, the fragmentation is not always to be a very good approximation[14,15,11,13].

It is worth while here to talk a little about the interesting and similar process, the production of $B_c$ and $B_c^*$ mesons in photon-photon collisions. A comparative study of the full $\alpha^2 \alpha_s^2$ calculation with the fragmentation approximation in the photon-photon process was presented in Ref.[16,17], where it was claimed that the fragmentation approximation is not valid. There are 20 Feynman diagrams which can be divided into four gauge invariant subsets corresponding to various attachments of photons onto the quark lines; i.e., subsets I, II, III, and IV corresponding respectively to the attachment of both photons onto the $b$ quark line, onto the $\bar{c}$ quark line, one photon onto the $b$ quark line and the other onto the $\bar{c}$, and the interchange of $b$ and $\bar{c}$. Subset I is dominated by the $\bar{b}$ quark fragmentation into the $B_c$ when $P_T$ is large, as discussed above. Subsets III and IV, the so-called recombination diagrams, can contribute to the non-fragmentation component substantially and decrease rapidly as $P_T$ is increasing, especially, when $P_T \gg M_{B_c}$. However, subset II is somewhat unusual. Although this contribution is relatively suppressed by the smaller probability for subsequent $b\bar{b}$ quark creation, it nevertheless gives quite a large contribution to the total cross section because of the enhancement of the $c$ quark electric charge. For instance, it has been found that when $\sqrt{\hat{s}} = 100$ GeV the result of the full calculation is an order of magnitude larger than the fragmentation calculation, even for large $P_T$. However, when $\sqrt{\hat{s}}$ becomes extremely large; e.g., $\sqrt{\hat{s}} = 800$ GeV, the contribution of this subset II is dominated by the $c$ quark fragmentating into the $B_c$ meson when $P_T \gg M_{B_c}$. This implies that naive power counting rules are violated[20]. However, such a kind of contribution in photon-photon collisions, which involves $b, \bar{b}$ quark pair creation, is not important in the hadronic productions of the $B_c$ and $B_c^*$, simply because such an enhancement factor of 16, the ratio of the electric charges of $c$ and $b$ quarks, does not occur at all.

The factors emphasized above may also be shown in the way: to present the cross sections for the hadronic productions of the $B_c$ and the $B_c^*$ mesons with various $P_T$ cuts, that is related to experiments more directly. Considering the situation of the experiments and experimental results at Tevatron being available soon, the cross sections at Tevatron energy and with various cuts for $P_T$ and a fixed one $|Y| < 1.5$ are calculated by the two approaches: the $\alpha_s^4$ calculation and the fragmentation approximation. The results are shown in Table I. Here the CTEQ3M parton distribution functions[19] is adopted.

**Table I. Total cross sections** $\sigma(P_{T\,B_c} > P_{T\,min})$ **at Tevatron in** $nb$

| $P_{T\,min}$ (GeV) | 0 | 5 | 10 | 15 | 20 | 25 | 30 |
|---|---|---|---|---|---|---|---|
| $\sigma_{B_c}(\alpha_s^4)$ | 1.8 | 0.57 | 0.087 | 0.018 | $4.8 \times 10^{-3}$ | $1.6 \times 10^{-3}$ | $6.3 \times 10^{-4}$ |
| $\sigma_{B_c}(frag.)$ | 1.4 | 0.47 | 0.071 | 0.014 | $4.0 \times 10^{-3}$ | $1.3 \times 10^{-3}$ | $5.3 \times 10^{-4}$ |
| $\sigma_{B_c^*}(\alpha_s^4)$ | 4.4 | 1.4 | 0.22 | 0.041 | $1.1 \times 10^{-2}$ | $3.4 \times 10^{-3}$ | $1.3 \times 10^{-3}$ |
| $\sigma_{B_c^*}(frag.)$ | 2.3 | 0.78 | 0.12 | 0.025 | $6.8 \times 10^{-3}$ | $2.3 \times 10^{-3}$ | $9.2 \times 10^{-4}$ |
| $\dfrac{\sigma_{B_c^*}(frag.)}{\sigma_{B_c^*}(\alpha_s^4)}$ | 0.52 | 0.55 | 0.56 | 0.61 | 0.63 | 0.67 | 0.70 |

We should note here that the values in the table are greater than those obtained in the Refs.[6,14] by a factor about two, it is due to more careful and reasonable considerations on the coupling constant $\alpha_s$ i.e. in the estimates for Table I, the coupling constant $\alpha_s(P_T^2)$ (with $P_T \gg m_b$) is adopted when it related to the part of the $b$ and $\bar{b}$ pair production (with a high $P_T$ in the considering mechanism), whereas $\alpha_s(4m_c^2)$ is adopted for the part of the fragmentation (the $c\bar{c}$ pair producion and the hadronization for the meson).

As pointed out by us[14,15], it is difficult to make a definite conclusion only from the $P_T$ distribution consideration if and where the fragmentation approximation is reliable for the hadronic productions of the $B_c$ and $B_c^*$ mesons. In fact, we have found that sometimes the consideration of $P_T$ alone can be misleading.

To clarify this issue, the newly introduced distribution $\sigma(z)$[15]:

$$\sigma(z) = \int dx_1 dx_2 g_1(x_1,\mu) g_2(x_2,\mu) \frac{d\hat{\sigma}(\sqrt{\hat{s}},\mu)}{dz}, \qquad (2)$$

is much more efficient. Here $z \equiv \dfrac{2(k_1+k_2)\cdot p}{\hat{s}}$ and $g_i(x_i,\mu)$ are the gluon distribution functions. In the subprocess center of mass $z$ is simply twice the fraction of the total energy carried by the $B_c$ or $B_c^*$ meson. The distribution $\sigma(z)$ provides a sensitive means to investigate the dynamics of the production processes and the fragmentation approximation. Clearly, if the fragmentation approximation is valid, $\dfrac{d\hat{\sigma}(\sqrt{\hat{s}},\mu)}{dz}$ can be factorized further as

$$\frac{d\hat{\sigma}(\sqrt{\hat{s}},\mu)}{dz} = \sum_i \hat{\sigma}_{gg \to Q_i \bar{Q}_i} \otimes D_{Q_i \to B_c}(z,\mu). \qquad (3)$$

where $D_{Q_i \to B_c}(z,\mu)$ are the usual fragmentation functions and $\hat{\sigma}_{gg \to Q_i \bar{Q}_i}$ is the gluon fusion subprocess cross section for the production of the heavy quark pair $Q_i \bar{Q}_i$. In the approximation, the integrals over $x_1$ and $x_2$ can be performed, and the fragmentation function can be factored out: $\sigma(z)$ is simply related to the fragmentation functions but not much sensitive to the parton distribution functions and the kinematic cuts as well. Therefore, comparing $\sigma(z)$ calculated in the fragmentation approximation with the full order $\alpha_s^4$ calculation, provides a quantitative criterion to judge the validity of the fragmentation approximation. Note that $z$ is a very useful variable and is an experimentally measurable quantity too.

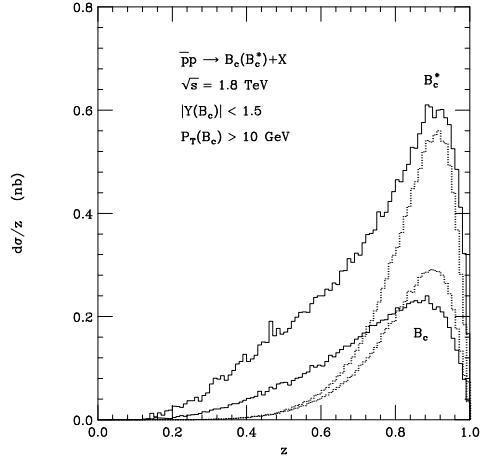

FIGURE 3a

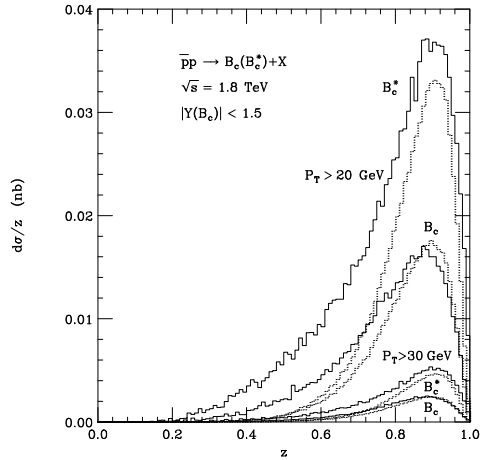

FIGURE 3b

Figure 3: $d\sigma/dz$ **versus** $z$ **at Tevatron**

In Fig 3, the two figures: Fig 3a with a cut $P_T > 10$ GeV and Fig 3b with cuts $P_T > 20$ GeV and $P_T > 30$ GeV are included, and the solid lines indicate those for the full QCD $\alpha_s^4$ calculations; the dotted lines indicate those for the fragmentation calculations. Since the cross sections for the $B_c$ and $B_c^*$ meson productions decrease very rapidly as $P_T$ increases, the distribution $\sigma(z)$ is sensitive to the smallest $P_T$ region through given $P_T$ cuts. From Fig. 3, some general features are evident: as for the $B_c$ meson, the distribution $\sigma(z)$ obtained by the fragmentation approximation is overestimated in the higher $z$ region while it is underestimated in the lower $z$ region for a small $P_T$ cut; but, after integration over $z$, the result is similar to the full $\alpha_s^4$ calculation, whereas, as for the $B_c^*$ meson, even for a large $P_T$ cut, the distribution $\sigma(z)$ is underestimated at all values of $z$ and, after integration over $z$, the result is definitely smaller than the full $\alpha_s^4$ calculation. This shows that it is simply fortuitous that the $P_T$ distribution of the $B_c$ calculated in the fragmentation approximation is similar to that from the full $\alpha_s^4$ calculation for $P_T$ below a certain value, particularly down to $P_T \sim M_{B_c}$. It is also clear that when $P_T$ is increased the distributions obtained by the two approaches become closer. As shown in Fig. 3b, when $P_T$ becomes as large as 30 GeV the curves calculated in the fragmentation approximation are quite close to the full $\alpha_s^4$ calculation. This indicates that the fragmentation approximation is valid in the large $P_T$ region and the higher the energy of the subprocess the higher the $P_T$ is requested. Perhaps the complicated feature pointed out above causes the authors of Refs.[11,13] to draw different conclusions. We should emphasize here one more point that the difference between the fragmentation approximation and full calculation is not universal, but is process-dependent, and it does not satisfy the Atarelli-Parisi evolution.

Finally, as discussed above, for $B_c^*$ meson production the fragmentation approach always underestimates the full $\alpha_s^4$. The deviation from the full calculation for the $B_c^*$ meson can be used as a criterion to test the validity of the fragmentation approximation. The results for the total cross section $\sigma(P_T > P_{T\min})$ for various $P_T$ cuts are listed in Table I. Taking agreement within 30% as the criterion for the validity of the fragmentation approximation from Table I, one may see that $P_T$ should exceed about 30 GeV at Tevatron, a value considerably much larger than the heavy quark masses. If comparing the ratio of $B_c^*$ to $B_c$ production predicted by the full $\alpha_s^4$ calculation with the fragmentation approximation, a similar conclusion may also be reached. We note that this conclusion is also rather insensitive to the choice of the QCD energy scale $\mu$ and the parton distribution functions.

### Acknowledgements

The author would like to thank his collaborators Yu-Qi Chen, Robert J. Oakes, Hong-Tao Jiang and Guo-Ping Han specially, because the talk is based on the pleasant collaboration. He would like to thank the Organizers of the INAUGURATION CONFERENCE OF APCTP for inviting him to talk in the conference and wishes APCTP very successful. He also would like to thank S. Randjbar-Daemi for the warm hospitality during his associate visit of ICTP. The work was supported in part by the National Natural Science Foundation of China and the Grant LWTZ-1298 of Chinese Academy of Sciences. The work was also supported in part by UNESCO and the International Atomic Energy Agency through ICTP.